\documentclass[twocolumn,superscriptaddress]{revtex4}

\newif\ifpdf\ifx\pdfichsageja\undefined\pdffalse\else\pdftrue\fi
\ifpdf\pdfoutput=1
\usepackage[pdftex]{graphicx}\usepackage{epstopdf}\pdfcompresslevel=9
\else\usepackage{graphicx}\fi                                            %

\usepackage[ansinew]{inputenc}
\usepackage[T1]{fontenc}

\usepackage{graphicx}
\usepackage{subfigure}
\usepackage{amsmath}
\usepackage{amssymb}
\usepackage{bbm}

\newcommand{\eps}{\varepsilon}
\newcommand{\kap}{\kappa}
\newcommand{\kaptil}{\tilde{\kappa}}
\newcommand{\halb}{\frac{1}{2}}
\newcommand{\eq}[1]{Eq.~(#1)}
\newcommand{\fig}[1]{Fig.~#1}

\newcommand{\mittel}[1]{\langle #1 \rangle}

\begin{document}

\date{December 2007]}
\title{Patterns of Chaos Synchronization}

\author{Johannes~Kestler}
\affiliation{Institute for Theoretical Physics, University of W\"urzburg, Am Hubland, 97074~W\"urzburg, Germany}
\author{Evi~Kopelowitz}
\affiliation{Department of Physics, Bar-Ilan University, Ramat-Gan, 52900 Israel}
\author{Ido~Kanter}
\affiliation{Department of Physics, Bar-Ilan University, Ramat-Gan, 52900 Israel}
\author{Wolfgang~Kinzel}
\affiliation{Institute for Theoretical Physics, University of W\"urzburg, Am Hubland, 97074~W\"urzburg, Germany}

\begin{abstract}
Small networks of chaotic units which are coupled by their time-delayed variables, are investigated. In spite of the time delay, the units can synchronize isochronally, i.e.\ without time shift. 
Moreover, networks can not only synchronize completely, but can also split into different synchronized sublattices. These synchronization patterns are stable attractors of the network dynamics. 
Different networks with their associated behaviors and synchronization patterns are presented.
In particular, we investigate sublattice synchronization, symmetry breaking, spreading chaotic motifs, synchronization by restoring symmetry and cooperative pairwise synchronization of a bipartite tree.
\end{abstract}

\maketitle

\section{Introduction}
Chaos synchronization is a counter-intuitive phenomenon. On one hand, a chaotic system is unpredictable. Two chaotic systems, starting from almost identical initial states, end in completely different trajectories. On the other hand, two identical chaotic units which are coupled to each other can synchronize to a common chaotic trajectory. The system is still chaotic, but after a transient the two chaotic trajectories are locked to each other \cite{Pikovsky:book, Schuster:2005}. This phenomenon has attracted a lot of research activities, partly because chaos synchronization has the potential to be applied for novel secure communication systems \cite{Pecora:1990, Cuomo:1993}. In fact, synchronization and bit exchange with chaotic semiconductor lasers has recently been demonstrated over a distance of 120 km in a public fiber-optic communication network \cite{Argyris:2005}. In this case, the coupling between the chaotic lasers was uni-directional, the sender was driving the receiver. For bi-directional couplings, when two chaotic units are interacting, additional interesting applications have been suggested. 
In this case, a secure communication over a public channel may be established. 
Although the algorithm as well as all the parameters are public, 
it is difficult for an attacker to decipher the secret message \cite{Klein:2006,Klein:2005,Kanter:Shutter:2007:PRL}. 

Typically, the coupling between chaotic units has a time delay due to the transmission of the exchanged signal. Nevertheless, chaotic units can synchronize without time shift, isochronically, although the delay time may be extremely long compared to the time scales of the chaotic units. 
This -- again counter-intuitive -- phenomenon has recently been demonstrated with chaotic semiconductor lasers \cite{Klein:06:PRE73, Fischer:2006:PRL, Sivaprakasam:2003, Lee:2006:JOSAB}, and it is discussed in the context of corresponding measurements on correlated neural activity \cite{Adrian:2006:SCI, Engel:1991:SCI, Campbell:1998, Rosenbluh:SpikingOpticalPatterns:2007:PRE}. 

Several chaotic units may be coupled to a network with delayed interactions. Such a network can synchronize completely to a single chaotic trajectory, or it may end in a state of several clusters, depending on the topology of the network or the distribution of delay times \cite{Atay:2004:PRL, Matskiv:2004, Masoller:2005, Topaj:2001}. Recently another phenomenon has been reported for chaotic networks: Sublattice synchronization. If a small network can be decomposed into two sublattices, then the units in each sublattice can synchronize to a common chaotic trajectory although they are not directly coupled. The coupling of one sublattice is relayed by the chaotic trajectory of a different sublattice. The trajectories of different sublattices are only weakly correlated, but not synchronized \cite{Sublattice:2007:PRE}. 

In this paper we want to investigate patterns of chaos synchronization for several lattices with uni- and bi-directional couplings with time delay. There exists a mathematical theory to classify possible solutions of nonlinear differential or difference equations for a given lattice \cite{Golubitsky:2006}. However, this theory does not determine the stability of these solutions. But in order to describe physical or biological dynamic networks, we are interested in stable patterns of chaotic networks. The patterns which are discussed in this paper are attractors in phase space, any perturbation of the system will relax to these patterns which move chaotically on some high dimensional synchronization manifold.

Our results are demonstrated for iterated maps, for the sake of simplicity and since we can calculate the stability of these networks analytically. But we found these patterns for other systems, as well, for example for the Lang-Kobayashi rate equations describing semiconductor lasers. Hence we think that our results are generic. 

\section{Two interacting units}
\label{sec:two}
\begin{figure} \centering 
	\vspace*{0.1cm}
	\includegraphics[scale=0.9]{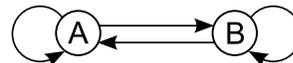}
	\vspace*{0.1cm}
	\caption{Two mutually coupled units.}
	\label{fig:zwei:setup}
\end{figure}

We start with the simplest network: Two units with delayed couplings and delayed self-feedback, as sketched in \fig{\ref{fig:zwei:setup}}.

For iterated maps, this network is described by the following equations:
\begin{equation} \label{eq:zwei:grundgleichungen}
\begin{split}
	a_t = (1 - \eps) f(a_{t - 1}) + \eps \kap f(a_{t - \tau}) + \eps (1 - \kap) f(b_{t - \tau}) \\
	b_t = (1 - \eps) f(b_{t - 1}) + \eps \kap f(b_{t - \tau}) + \eps (1 - \kap) f(a_{t - \tau}) 
\end{split}
\end{equation}
where $f(x)$ is some chaotic map, for example the Bernoulli shift,
\begin{equation}
	f(x) = \alpha \, x \mod 1
\end{equation}
with $\alpha > 1$. In this case, the system is chaotic for all parameters $0 < \eps < 1$ and $0 < \kap < 1$. $\eps$ measures the total strength of the delay terms and $\kap$ the strength of the self-feedback relative to the delayed coupling. 

\begin{figure} \centering 
	\includegraphics[scale=0.35]{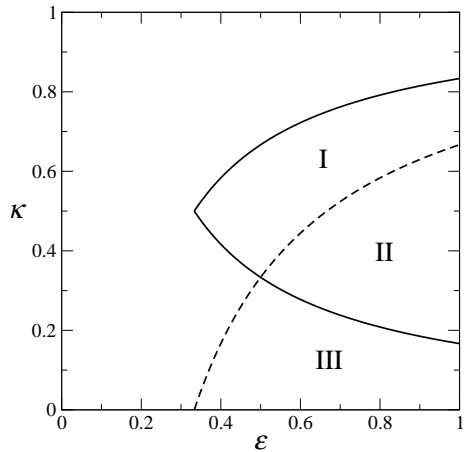}
	\caption{Phase diagram for $\alpha = 3/2 \,$ (analytical result).}
	\label{fig:zwei:gebiete}
\end{figure}

Obviously, the synchronized chaotic trajectory $a_t = b_t$ is a solution of \eq{\ref{eq:zwei:grundgleichungen}}. Its stability is determined by $\tau$ conditional Lyapunov exponents which describe the stability of perturbations perpendicular to the synchronization manifold. For the Bernoulli map, these Lyapunov exponents have been calculated analytically \cite{Lepri:1993:PHD, Sublattice:2007:PRE}, and for infinitely long delay, $\tau \to \infty$, one obtains the phase diagram of \fig{\ref{fig:zwei:gebiete}}.

In region I and II, i.e.\ for
\begin{equation}
\frac{\alpha - 1}{2 \alpha \eps} < \kap < \frac{2 \alpha \eps + 1 - \alpha}{2 \alpha \eps},
\label{for:zwei:gebietAB}
\end{equation}
the two units are synchronized to an identical chaotic trajectory $a_t = b_t$. Although the two units are coupled with a long delay $\tau$, they are completely synchronized without any time shift. 
For $\tau \to \infty$, this region is symmetric about the line $\kap = \halb$. 

Complete synchronization can be understood by considering a single unit driven by some signal $s_t$:
\begin{equation} \label{eq:zwei:getrieben}
	a_t = (1 - \eps) f(a_{t - 1}) + \eps \kaptil f(a_{t - \tau}) + s_{t-\tau} \,.
\end{equation} 
If the system is not chaotic, i.e.\ if its Lyapunov exponents are negative, then the trajectory $a_t$ relaxes to a unique trajectory determined by the drive $s_t$. For the Bernoulli shift, this region is defined by the inequality
\begin{equation}
\left( -\frac{1 + \alpha \eps - \alpha}{\alpha \eps} < \right) \kaptil < \frac{1 + \alpha \eps - \alpha}{\alpha \eps}
\label{for:zwei:gebietMSS}
\end{equation} 
and indicated by II + III in \fig{\ref{fig:zwei:gebiete}}. Since the trajectory which A relaxes to is uniquely determined by the drive, further identical units receiving the same drive would relax to the same trajectory. Thus \eq{\ref{for:zwei:gebietMSS}} defines the region where identical units which get the same input synchronize to a common trajectory.

Now let us rewrite \eq{\ref{eq:zwei:grundgleichungen}}:
\begin{equation} \label{eq:umgeschrieben}
\begin{split}
	&a_t = (1 - \eps) f(a_{t - 1}) + \eps (2 \kap - 1) f(a_{t - \tau}) + {} \\
		& \qquad\quad {} + \eps (1 - \kap) f(a_{t - \tau}) + \eps (1 - \kap) f(b_{t - \tau}) \, ,\\
	&b_t = (1 - \eps) f(b_{t - 1}) + \eps (2 \kap - 1) f(b_{t - \tau}) + {} \\
		& \qquad\quad {} + \eps (1 - \kap) f(b_{t - \tau}) + \eps (1 - \kap) f(a_{t - \tau}) .
\end{split}
\end{equation} 
Both systems are driven by the identical signal
\begin{equation}
	s_{t-\tau} = \eps (1 - \kap) \big[ f(a_{t - \tau}) + f(b_{t - \tau}) \big] .
\end{equation} 
Comparing (\ref{eq:zwei:getrieben}) with (\ref{eq:umgeschrieben}), one finds that for
\begin{equation} \label{eq:zwei:skalierung}
	\kaptil = 2 \kap - 1
\end{equation} 
the interacting units (\ref{eq:umgeschrieben}) are described by the driven unit (\ref{eq:zwei:getrieben}).
The phase boundary of the driven system, region II + III [\eq{\ref{for:zwei:gebietMSS}}],
and the phase boundary of the interacting system, region I + II [\eq{\ref{for:zwei:gebietAB}}],
are connected with each other: With the mapping of \eq{\ref{eq:zwei:skalierung}}, one phase boundary can be obtained from the other. This is not only true for $\tau \to \infty$ but for any value of $\tau$.
 
Additionally, this mapping does not only hold for the Bernoulli shift but for any chaotic system, provided that the signal does not change the Lyapunov exponents of the driven system. Simulations with other maps, e.g.\ with the Tent map or the Logistic map, confirm the applicability of the mapping (\ref{eq:zwei:skalierung}). Since the slopes of these maps are not constant, the Lyapunov exponents depend on the trajectories; in order to get trajectories which are comparable to the trajectories of two mutually coupled units, the driving signal 
in the driver/receiver setup
must come from an identical unit $c_t$ (which has an increased self-coupling due to the lack of external couplings), i.e.
\begin{eqnarray}
	s_{t-\tau} & = & \eps (1 - \kaptil) f(c_{t-\tau}) \quad\text{with} \\
	c_t & = & (1 - \eps) f(c_{t-1}) + \eps f(c_{t-\tau}) \, .
\end{eqnarray} 

Even for the Lang-Kobayashi equations \cite{Lang:1980,Ahlers:1998}, which describe coupled semiconductor lasers, we found a good agreement for the mapping (\ref{eq:zwei:skalierung}), see section \ref{sec:lk}. 

Now let us come back to the iterated equations (\ref{eq:zwei:grundgleichungen}). Let us assume that we record the synchronized trajectory $a_t = b_t$ of two interacting chaotic units, regions I and II of \fig{\ref{fig:zwei:gebiete}}. Now we insert the recorded trajectory $b_t$ into \eq{\ref{eq:zwei:grundgleichungen}}. How will $a_t$ respond to this drive? We find that in 
region II
the unit A will synchronize completely to the recorded trajectory $b_t$, whereas in region I the unit A does not synchronize. Although the two interacting units A and B do synchronize, the unit A does not follow the recorded trajectory in region I. This shows that bi-directional interaction is different from uni-directional drive.

\section{Sublattice synchronization}
\label{sec:sublattice}

\begin{figure} \centering 
	\includegraphics[scale=0.9]{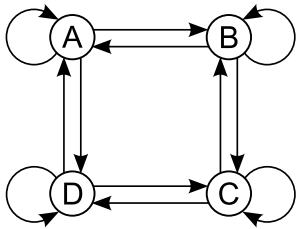}
	\caption{Ring of 4 units.}
	\label{fig:4erRing:setup}
\end{figure}

The response of a single chaotic unit to an external drive, \fig{\ref{fig:zwei:gebiete}}, determines also the phase diagram of a ring of four chaotic units. Additionally, it shows a new phenomenon: sublattice synchronization \cite{Sublattice:2007:PRE}. Consider the ring of four identical units of \fig{\ref{fig:4erRing:setup}}. The dynamics of unit A is described by
\begin{multline}
	a_t = (1 - \eps) f(a_{t - 1}) + \eps \kap f(a_{t - \tau}) + {} \\
	{} + \eps (1 - \kap) \left[ \halb f(b_{t - \tau}) + \halb f(d_{t - \tau}) \right] ;
\end{multline} 
the dynamics of B, C and D are defined analogously. Note the additional weighting $\halb$ for the external coupling (to 2 neighbors) which causes the total strength of the external coupling to remain $\eps (1 - \kap)$.

Obviously, the two units A and C receive identical input from the units B and D. Consequently, they will respond with an identical trajectory in the regions II and III of \fig{\ref{fig:zwei:gebiete}}.
The same argument holds for the two units B and D,
which both get the same input from A and C.
That leads to sublattice synchronization in region III of \fig{\ref{fig:zwei:gebiete}}: A and C have an identical chaotic trajectory and B and D have a different one. Although there is a delay of arbitrary long time of the transmitted signal, synchronization is complete, without any time shift. The synchronization of A and C is mediated by the chaotic trajectory of B and D and vice versa. But the two trajectories have only weak correlations, they are not synchronized. Numerical simulations of the Bernoulli system with small values of $\tau$ have shown that there is no generalized synchronization either \cite{Abarbanel:1996:PRE,Boccaletti:2002:PRep,Sublattice:2007:PRE}.

\begin{figure} \centering 
	\includegraphics[scale=0.5]{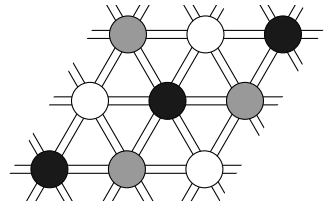}
	\caption{Sublattice synchronization in a triangular lattice with periodic boundaries. The double lines signify bi-directional couplings. The self-feedback is not drawn to simplify the illustration.}
	\label{fig:Dreiecksgitter:setup}
\end{figure}

Sublattice synchronization has been found for 
rings with an even number of units, for chains and also for
other lattices
which can be decomposed into identical sublattices \cite{Sublattice:2007:PRE}.
For example, the lattice of \fig{\ref{fig:Dreiecksgitter:setup}} can be decomposed into three sublattices. 
For some parameters of the Bernoulli system we find sublattice synchronization with three chaotic trajectories. Again, the synchronized units are not directly coupled, but they are indirectly connected via the trajectories of the other sublattices. 

The sublattice trajectories, described before, are stable, i.e.\ the conditional Lyapunov exponents which describe perturbations perpendicular to the synchronization manifold are negative. Even when the system starts from random initial states it relaxes to the state of sublattice synchronization (region III of \fig{\ref{fig:zwei:gebiete}} for the ring of 4 units).
The chaotic trajectories of sublattice synchronization may be depicted as 
$\begin{smallmatrix} \text{A} & \text{B} \\ \text{B} & \text{A} \end{smallmatrix}$.
Note that this structure does not break the symmetry of the ring: The statistical properties of the chaotic trajectory of A is identical to those of B.

\begin{figure} \centering 
	\subfigure[Two different coupling parameters $\kap_1$ and $\kap_2$.]{
		\includegraphics[scale=0.9]{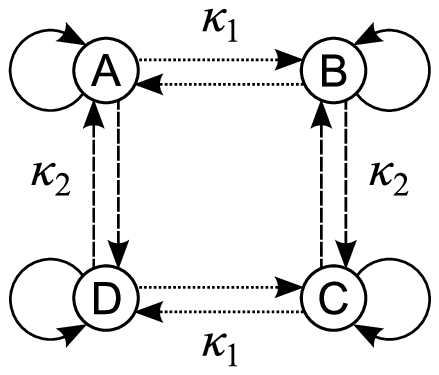}
	}
	\qquad
	\subfigure[Phase diagram for Bernoulli system with $\alpha = 3/2$ and $\eps = 11/20$ (analytical result). The regions of no synchronisation are labeled with a dash ($-$).
	]{
		\includegraphics[scale=0.35]{fig5b.eps}
	}
	\caption{Ring of 4 units with two different coupling parameters.}
	\label{fig:zweiKopplungen}
\end{figure}

However, there are other solutions of the dynamic equations, as well. These solutions are classified according to the theory of Golubitsky et al \cite{Golubitsky:2006}. For example, for the ring, the state 
$\begin{smallmatrix} \text{A} & \text{A} \\ \text{B} & \text{B} \end{smallmatrix}$
is a solution as well. Such a state breaks the symmetry of the lattice. But we find that this state is unstable. Any tiny perturbation relaxes to the states
$\begin{smallmatrix} \text{A} & \text{B} \\ \text{B} & \text{A} \end{smallmatrix}$
in region III,
$\begin{smallmatrix} \text{A} & \text{A} \\ \text{A} & \text{A} \end{smallmatrix}$
in region II and
$\begin{smallmatrix} \text{A} & \text{B} \\ \text{D} & \text{C} \end{smallmatrix}$
outside of II and III. In fact, we have never found a stable state which breaks the symmetry of the lattice. Only when the couplings in the two directions of the ring are different, the states
$\begin{smallmatrix} \text{A} & \text{A} \\ \text{B} & \text{B} \end{smallmatrix}$
and
$\begin{smallmatrix} \text{A} & \text{B} \\ \text{A} & \text{B} \end{smallmatrix}$
are stable in some (different) parts of the parameter space, see \fig{\ref{fig:zweiKopplungen}}. 
Hence we postulate that spontaneous symmetry breaking is not possible for finite lattices of chaotic units.

\section{Spreading chaotic motifs}
\label{sec:motifs}

\begin{figure} \centering 
	\subfigure[Synchronization pattern (Sublattice synchronization) of region III.]{
		\includegraphics[scale=0.55]{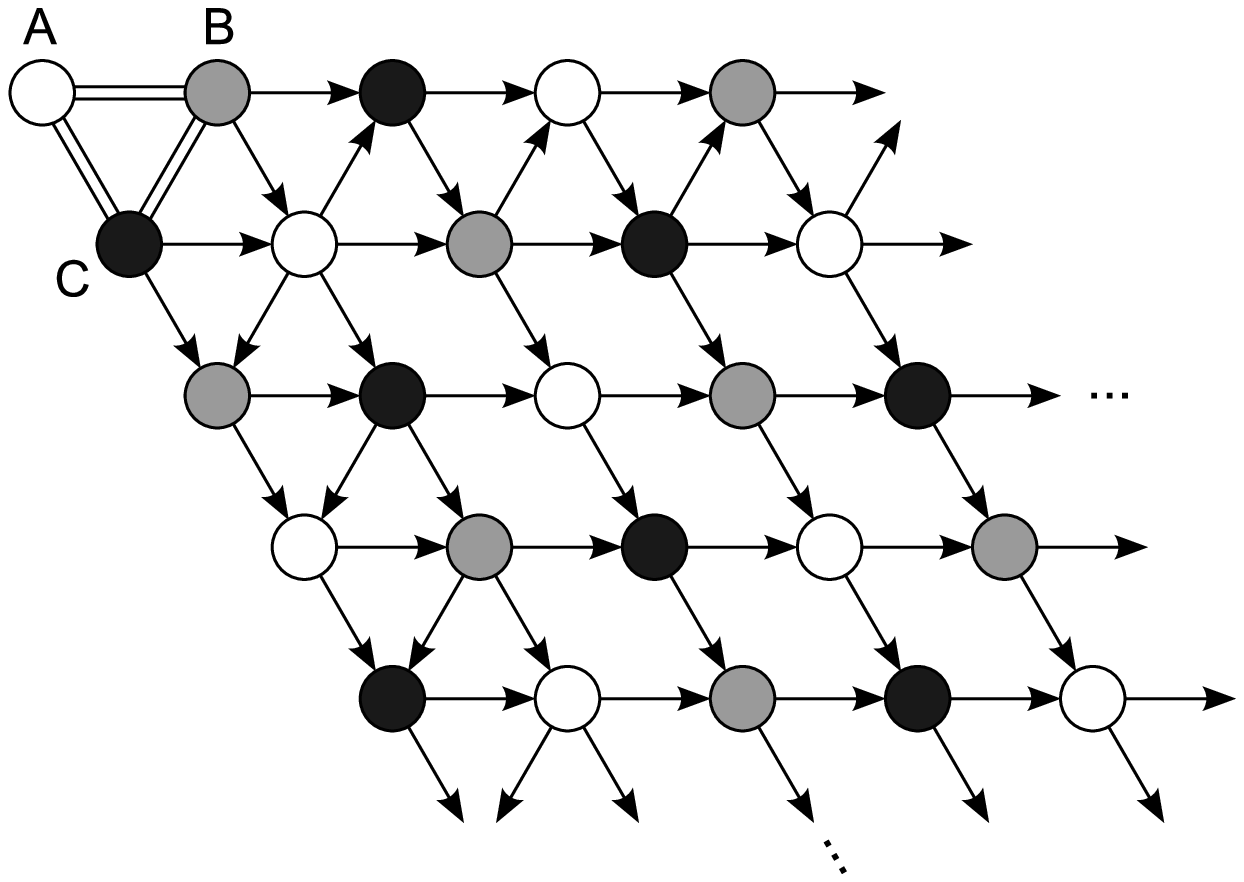}
		\label{fig:2dim:setup}
	}
	\quad
	\subfigure[Phase diagram for Bernoulli system, $\alpha = 3/2$ . Analytical result combining the analytical synchronization region for a ring of 3 units (---) and the region where identical units which are driven by an identical signal, synchronize (- -).]{
		\includegraphics[scale=0.35]{fig6b.eps}
		\label{fig:2dim:gebiete}
	}
	\caption{Triangle (three bi-directionally coupled units) with a uni-directionally attached infinitely large lattice. Double lines signify bi-directional couplings whereas arrows show uni-directional couplings. The self-feedback is not drawn to simplify the illustration.}
	\label{fig:2dim:gebiete:ab}
\end{figure}

The response of a chaotic unit to an external drive, \fig{\ref{fig:zwei:gebiete}}, points to another interesting phenomenon. Consider a triangle of chaotic units with bi-directional couplings as sketched in \fig{\ref{fig:2dim:setup}}.
Choose the parameters such that the triangle is completely disordered, but each unit has negative Lyapunov exponents when it is separated from the two others. (Both conditions are fulfilled in region III of \fig{\ref{fig:2dim:gebiete}}, which shows analytical results for the Bernoulli system.) When we record the three time series $a_t$, $b_t$ and $c_t$ we find three different weakly correlated chaotic trajectories. Now we feed the two trajectories $b_t$ and $c_t$ into an infinitely large lattice of identical units with uni-directional couplings as shown in \fig{\ref{fig:2dim:setup}}. Each unit receives two input signals from two other units. But since all Lyapunov exponents are negative, 
the system responds with the three chaotic trajectories $a_t$, $b_t$ and $c_t$. Although the units of the initial triangle are not synchronized, their pattern of chaos is transmitted to the infinite lattice.
All units of the same sublattice are completely synchronized without time shift, although the coupling has a long delay time $\tau$. 
Hence the chaotic motif, three weakly correlated chaotic trajectories, can be imposed on an arbitrarily large lattice. Note that the time for spreading a motif on a large lattice increases only linearly with the number of units because of the unidirectional couplings.

For some parameters $\kap$ and $\eps$, namely in regions I and II of \fig{\ref{fig:2dim:gebiete}}, the three units of the triangle are completely synchronized. In region I, only the three units are synchronized while the other units remain un-synchronized. In region II, the other units, too, get synchronized to the triangle (because all Lyapunov exponents are negative), so the whole lattice is completely synchronized.

The phenomenon of spreading motifs is not restricted to a triangle. For example, a ring of 6 mutually coupled units has also a region in the corresponding $(\eps, \kap)$ phase diagram where there is no synchronization, but where isolated units have negative Lyapunov exponents, comparable to region III in \fig{\ref{fig:2dim:gebiete}} \cite{Sublattice:2007:PRE}. Consequently, the pattern of six chaotic trajectories can lead to sublattice synchronization with 6 different sublattices on a corresponding lattice of chaotic units with short-range unidirectional couplings.

\section{Synchronization by restoring symmetry}
\label{sec:symmetry}

In general we expect that the larger the network is, the smaller the region in the parameter space is where the network synchronizes. For example, a ring of $N = 6$ units has a smaller region of synchronization than the region II and III of \fig{\ref{fig:zwei:gebiete}} for $N = 4$. With increasing $N$ (and the slope $\alpha$ being held constant) synchronization finally disappears completely \cite{Sublattice:2007:PRE}. However, we found a counterexample where adding a unit restores synchronization. 

\begin{figure} \centering 
	\subfigure[Chain of five units with two different time delays and without self-feedback.]{
		\includegraphics[scale=0.8]{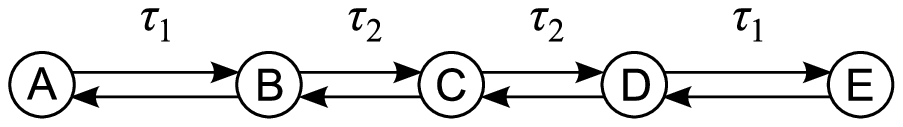}
		\label{fig:kette:setupmit}
	}
	\subfigure[The last unit has been removed.]{
		\includegraphics[scale=0.8]{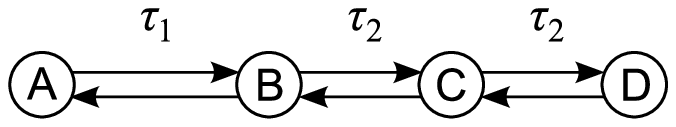}
		\label{fig:kette:setupohne}
	}
	\caption{Symmetric and asymmetric chain without self-feedback.}
	\label{fig:kette:setup}
\end{figure}

Consider the chain of 5 units shown in \fig{\ref{fig:kette:setupmit}}.
The coupling to the two outer units has a longer delay time than the internal couplings. There is no self-feedback, $\kap = 0$. 
Now remove unit E, \fig{\ref{fig:kette:setupohne}}, and rescale the coupling to unit D. (Unit D doubles the input from unit C to compensate for the missing second neighbor.) In this case, a synchronized solution does not exist. Numerical simulations of the Bernoulli system and the laser equations show high correlations between units A and C with time shift $\Delta=\tau_1-\tau_2$, and between B and D with zero time shift, but the correlation coefficient does not achieve the value one. On the other side, if we add unit E we restore the symmetry of the chain. In this case we find sublattice synchronization with time shift between the outer units and the central one:
\begin{equation}
	a_t = e_t = c_{t - \Delta}; \quad b_t = d_t \; .
\end{equation} 
If $\tau_1$ is greater than $\tau_2$, the central unit is earlier than the chaotic trajectory of the outer ones, it leads, whereas for the opposite case it lags behind. 

\section{Cooperative pairwise synchronization}
\label{sec:tree}

\begin{figure} \centering 
	\subfigure[There are $N$ different delay times which are pairwise identical for one unit of side A and one unit of side B.]{
		\includegraphics[scale=0.7]{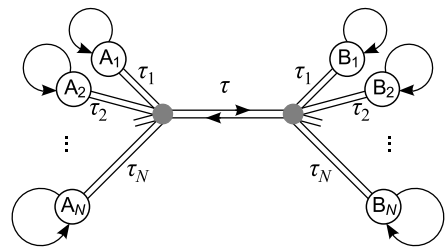}
		\label{fig:einKanal:setup}
	}
	\subfigure[There are $N$ different shifts in the Bernoulli map which are pairwise identical for one unit of side A and one unit of side B.]{
		\includegraphics[scale=0.7]{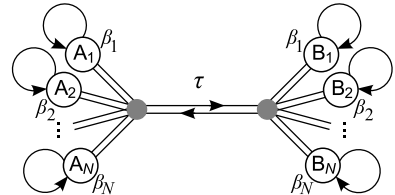}
		\label{fig:einKanal:setupShift}
	}
	\caption{Each unit on one side is coupled to all units of the other side.}
	\label{fig:einKanal:setupBeide}
\end{figure}

Is it possible to synchronize two sets of chaotic units with a single coupling channel? In fact, we found such examples where two sets of chaotic units are bi-directionally connected by the sum of their units, as indicated in \fig{\ref{fig:einKanal:setupBeide}}.
There are $2 N$ units, i.e.\ the number of units on each side is $N$. Each side is the mirror image of the other. 
Only one single bi-directional signal 
composed of all $N$ signals from each side
is driving the other side, 
and this leads to cooperative pairwise synchronization.

In the first setup, \fig{\ref{fig:einKanal:setup}}, all units are identical, but the delay times of their couplings are different. The units have pairwise identical delay times, i.e.\ $\text{A}_k$ and $\text{B}_k$ have a coupling delay time $2\tau_k+\tau$ which is enforced by a self-feedback with delay time $\tau_s=2\tau_k+\tau$. Hence, for one pair, $N = 1$,  we obtain the phase diagram of \fig{\ref{fig:zwei:gebiete}}, where the two units are completely synchronized in regions I and II. For a large number $N$ of units, each A unit receives the identical signal
\begin{equation}
	s_t = \eps (1 - \kap) \frac{1}{N} \sum_{k = 1}^N f(b_{k,t-(2\tau_k + \tau)})
\end{equation} 
and vice versa. Note that the unit $\text{A}_k$ receives only a weak signal of the order $1/N$ from its counterpart $\text{B}_k$. Nevertheless, we find that the network synchronizes to a state of pairwise identical chaotic trajectories, $a_{k,t}=b_{k,t}; k=1, \ldots, N$. For the Bernoulli system, the region of pairwise synchronization is similar to region II of \fig{\ref{fig:zwei:gebiete}}. There is no synchronization among units of the same side. Each unit receives the sum of all chaotic trajectories, but it responds only to the tiny part which belongs to its counterpart. The synchronization is a cooperative effect. As soon as a single unit is detuned, the whole network loses synchronization. 

The second setup, \fig{\ref{fig:einKanal:setupShift}}, is similar to the first one but allows analytical calculations. In this second setup, all delay times are identical, while each unit of one side has a different shift $\beta_k$, $k = 1, \ldots, N$, in its shifted Bernoulli map, see \eq{\ref{eq:shiftedBernoulli}}; the shifts are pairwise identical for unit A$_k$ and the corresponding unit B$_k$ so one side is again the mirror image of the other side. The shifted Bernoulli map is defined by
\begin{equation} \label{eq:shiftedBernoulli}
	f_{\beta}(x) = [\alpha (x + \beta)] \mod 1 \, .
\end{equation} 
Each A unit receives the signal
\begin{equation}
	s_t = \eps (1 - \kap) \frac{1}{N} \sum_{k = 1}^N f_{\beta _k}(b_{k,t - \tau})
\end{equation} 
and vice versa. Analytical calculations are possible for this second setup because the delay times are equal -- in contrast to the first setup -- and the different shifts in the map do not hamper the calculations.
If the Bernoulli maps were not shifted, the units would all be identical. Then in regions II and III of \fig{\ref{fig:zwei:gebiete}}, all units of the same side would synchronize, $a_{1,t} = a_{2,t} = \ldots = a_{N,t}$; $b_{1,t} = b_{2,t} = \ldots = b_{N,t}$ because they receive the same input. Then one would effectively get two coupled units A and B (which synchronize in regions I and II), so in region II all $2 N$ units would be synchronized. 
For \emph{shifted} Bernoulli maps, the units of the same side are not allowed to synchronize by definition. Nevertheless, the stability analysis regarding the linearized equations is not affected by the different shifts, meaning that the same perturbations still are damped in the same regions; the only difference is that due to the shifts, most of the trajectories are not allowed to come close together. Only the pairs of corresponding units can synchronize, $a_{k,t} = b_{k,t}$, and so they do in region II.

\section{Analogy of laser equations to iterated maps}
\label{sec:lk}

Besides iterated maps, we considered the Lang-Kobayashi equations, which describe the dynamics of semiconductor lasers optically coupled to their own or/and to the light of other semiconductor lasers. We used them for simulations in the following form, according to Ref.\ \cite{Ahlers:1998}:
\begin{multline}
\frac{d}{dt} E_{0,j}(t) = \halb G_N n_j(t) E_{0,j}(t) + \frac{C_{\text{sp}} \gamma [N_{\text{Sol}} + n_j(t)]}{2 E_{0,j}(t)}  \\
{} + \lambda \, E_{0,j}(t - \tau) \cos[\omega_0 \tau + \phi_j(t) - \phi_j(t - \tau)]  \\
{} + \sigma \sum_{k=1}^{N_{\text{lasers}}} \!\! w_{j,k} E_{0,k}(t - \tau) \cos[\omega_0 \tau + \phi_j(t) - \phi_k(t - \tau)] \, ,
\end{multline} 
\begin{multline}
\frac{d}{dt} \phi_j(t) = \halb \alpha_{\text{lef}} G_N n_j(t)  \\
{} - \lambda \, \frac{E_{0,j}(t - \tau)}{E_{0,j}(t)} \sin[\omega_0 \tau + \phi_j(t) - \phi_j(t - \tau)]  \\
{} - \sigma \sum_{k=1}^{N_{\text{lasers}}} \!\! w_{j,k} \frac{E_{0,k}(t - \tau)}{E_{0,j}(t)} \sin[\omega_0 \tau + \phi_j(t) - \phi_k(t - \tau)] \, ,
\end{multline} 
\begin{multline}
\frac{d}{dt} n_j(t) = (p - 1) \gamma N_{\text{sol}} - \gamma n_j(t) - [\Gamma + G_N n_j(t)] E_{0,j}^2(t) \, ,
\end{multline}
where $E_{0,j}(t)$ and $\phi_j(t)$  are the amplitude and the slowly varying phase of the electric field $E_j(t) = E_{0,j}(t) \exp\{\text{i}[\omega_0 \tau + \phi_j(t)]\}$ and $n_j(t)$ is the carrier number above the value for a solitary laser, $j = 1, \ldots, N_{\text{lasers}}$. The strength of the self-feedback is determined by $\lambda$, while the strength of the external coupling is defined by $\sigma$. The equations above cover the general case of $N_{\text{lasers}}$ coupled semiconductor lasers, where the network structure and coupling strengths are determined by the weightings $w_{j,k}$, which are normalized so that the strength of the total input for each laser is the same, $\sum_{k=1}^{N_{\text{lasers}}} w_{j,k} = 1$ for each laser $j$ \footnote{
Usually $w_{j,j} = 0$ because the self-feedback is already taken into account by the term weighted by $\lambda$. Only if a laser $j$ gets no input from other lasers, then $w_{j,j} = 1$ to compensate for the missing external input. (Alternatively one could increase the corresponding $\lambda$.)}.
The parameters $N_{\text{sol}}$, $G_N$, $\tau$, $\alpha_{\text{lef}}$, $\gamma$, $\Gamma$, $p$, $\omega_0$ and $C_{\text{sp}}$ are chosen according to Ref.\ \cite{Ahlers:1998}.

In a very simplified form, the equations above read
\begin{multline}
	\frac{d}{d t} x = [\text{internal dynamics}] + \lambda [\text{self-feedback}]  \\
	{} + \sigma [\text{external coupling}] \, .
	\label{eq:lkgleichung}
\end{multline}
Comparing equations (\ref{eq:zwei:grundgleichungen}) and (\ref{eq:lkgleichung}) yields a relation between the parameter space of the Lang-Kobayashi equations, $\{\lambda, \sigma\}$, and the parameter space of the maps, $\{\eps, \kap\}$, if the two following conditions are considered: 
\begin{itemize}
\item[(i)] 
The ratio of the self-feedback to the external coupling should be the same in both cases.
\item[(ii)]
The ratio of the time-delayed terms to the internal dynamics should be the same in both cases.
\end{itemize}
These two conditions yield the following transformation:
\begin{equation}
	\label{eq:transformation}
	\kap = \frac{\lambda}{\lambda + \sigma}, \quad \eps = \frac{\lambda + \sigma}{v} \, .
\end{equation} 
The second condition (ii) is not properly defined because the Lang-Kobayashi equations (\ref{eq:lkgleichung}) -- in contrast to the iterated equations (\ref{eq:zwei:grundgleichungen}) -- are differential equations and the internal dynamics is not only the first term on the right-hand side of \eq{\ref{eq:lkgleichung}} but is also contained in the current state $x$. Therefore the denominator $v$ in (\ref{eq:transformation}) is not given and has to be chosen reasonably. We took $v = 180 \text{ ns}^{-1}$ so $\eps$ is between $0$ and $1$ [because the sum $\lambda + \sigma$ (the strength of the re-injected light) should not exceed the value of $v = 180\text{ ns}^{-1}$ in our case].

In order to measure synchronization, we averaged the amplitudes of the electric fields, $E_{0,j}$, in $1$ ns intervalls and calculated the un-shifted, $\Delta = 0$, cross correlation function defined by
\begin{equation}
	C_{xy}^{\Delta} = \frac{\mittel{x_t \cdot y_{t-\Delta}} - \mittel{x} \mittel{y}}
		{\sqrt{\mittel{x^2} - \mittel{x}^2} \;\cdot \sqrt{\mittel{y^2} - \mittel{y}^2} \;} \, .
\end{equation} 
A cross correlation over $0.99$ can be regarded as synchronization in the case of lasers.

\begin{figure*}
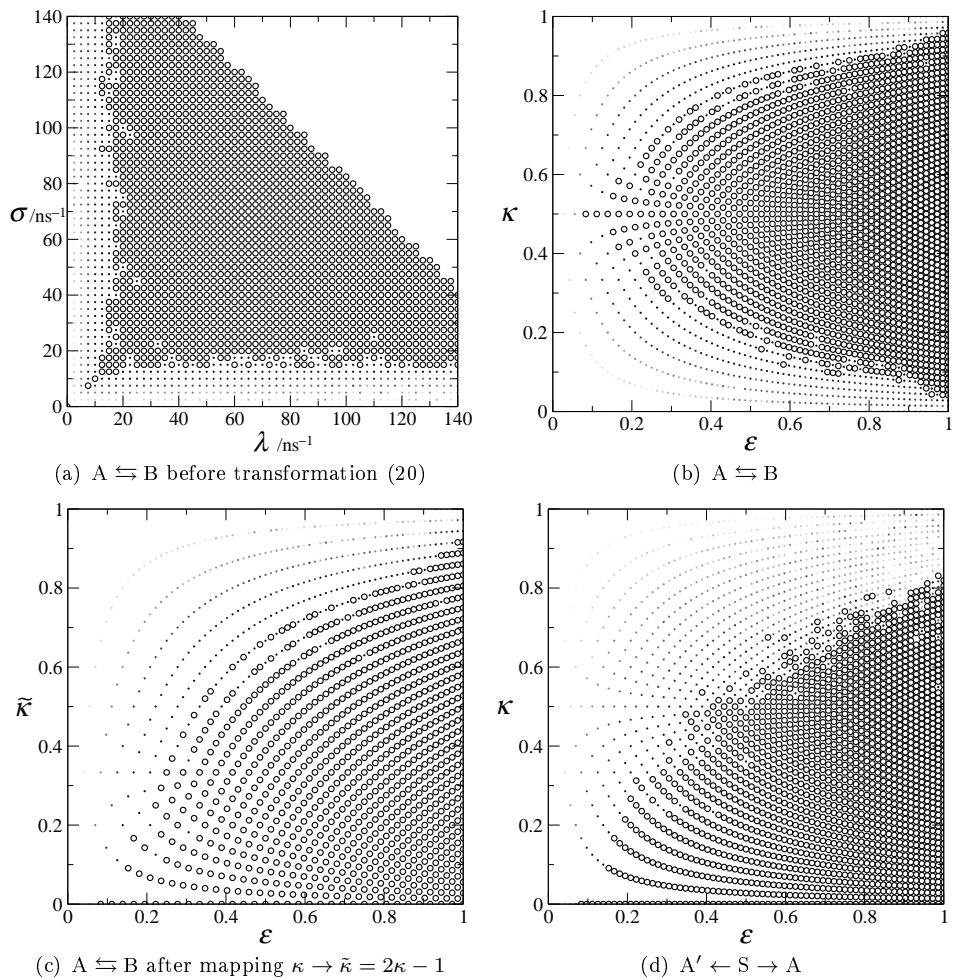
 \centering 
	\subfigure[$\text{A} \leftrightarrows \text{B}$ before transformation (\ref{eq:transformation})]{
		\includegraphics[height=0.25\textheight]{fig9a.eps}
	}
	\subfigure[$\text{A} \leftrightarrows \text{B}$]{
		\includegraphics[height=0.25\textheight]{fig9b.eps}
	}
	\subfigure[$\text{A} \leftrightarrows \text{B}$ after mapping $\kap \to \kaptil = 2 \kap - 1$]{
		\includegraphics[height=0.25\textheight]{fig9c.eps}
	}
	\subfigure[$\text{A}' \leftarrow \text{S} \rightarrow \text{A}$]{
		\includegraphics[height=0.25\textheight]{fig9d.eps}
	}
	\caption{Phase diagrams for the Lang-Kobayashi laser equations. Every circle/point represents one simulation. Open circles ($\circ$) show a cross correlation $C > 0.99$, which can be regarded as synchronization in the case of the laser equations. Figure (a) shows results for two mutually coupled lasers before applying transformation (\ref{eq:transformation}), wheras (b) shows the same diagram after the parameter transformation. The synchronization region of two mutually coupled units (b) can be mapped [using \eq{\ref{eq:zwei:skalierung}}] to a region (c) which is similar to the synchronization region of a driver/receiver system (d). After the parameter transformation (\ref{eq:transformation}), the synchronization regions for the Lang-Kobayashi equations, figures (b), (c) and (d), look very similar to the ones for the Bernoulli maps, \fig{\ref{fig:zwei:gebiete}}.}
	\label{fig:laser}
\end{figure*}

Figure \ref{fig:laser} shows both the parameter transformation, \eq{\ref{eq:transformation}}, from $\{\lambda, \sigma\}$ to $\{\eps, \kap\}$ and the mapping, \eq{\ref{eq:zwei:skalierung}}, from two mutually coupled units to driven ones. From the comparison of this figure with \fig{\ref{fig:zwei:gebiete}}, the analogy of laser equations to iterated maps can be seen. 
Besides the analogy of features described in section II, one can also find this analogy for the features described in sections \ref{sec:sublattice}, \ref{sec:symmetry} and \ref{sec:tree}.

\section{Summary}

Small networks of chaotic units with time-delayed couplings show interesting patterns of chaos synchrony. These patterns are stable attractors of the network dynamics.

The phase diagram of two and four units with mutual couplings has been related to the properties of a single driven chaotic unit.
Two interacting units with self-feedback can synchronize completely, without time shift, even if the delay time is extremely large.
When the chaotic trajectory of two interacting units is recorded and used to drive a single identical unit, it turns out that the driven unit synchonizes only in a small part of the phase diagram. Hence interaction is different from drive.

Sublattice synchronization is found for lattices which can be decomposed into a few sublattices. Each sublattice is completely synchronized, but different sublattices are only weakly correlated. Synchronization is relayed by different chaotic trajectories. The trajectories of each sublattice have identical statistical properties. Thus the symmetry of the lattice is not broken. There are solutions of the dynamic equations which break the symmetry of the lattice. However, we always found that these solutions are unstable. Hence we postulate that stable patterns of chaos synchrony possess the symmetry of the corresponding lattice.

Synchronization depends on the symmetry of the network. When the symmetry of an asymmetric chain is restored by adding units, sublattice or complete synchronization is restored, too.

Finally, a bi-partite network, where the two parts are coupled by a single mutual signal, shows pairwise complete synchronization, whereas the units of each part do not synchronize. Each unit responds to the weak contribution of its partner in the other part of the network. Pairwise synchronization is a cooperative effect: Detuning a single unit destroys the synchronization of the whole network.

The work of Ido Kanter is partially supported by the Israel Science Foundation.

\bibliography{patterns}

\begin{thebibliography}{27}
\expandafter\ifx\csname natexlab\endcsname\relax\def\natexlab#1{#1}\fi
\expandafter\ifx\csname bibnamefont\endcsname\relax
  \def\bibnamefont#1{#1}\fi
\expandafter\ifx\csname bibfnamefont\endcsname\relax
  \def\bibfnamefont#1{#1}\fi
\expandafter\ifx\csname citenamefont\endcsname\relax
  \def\citenamefont#1{#1}\fi
\expandafter\ifx\csname url\endcsname\relax
  \def\url#1{\texttt{#1}}\fi
\expandafter\ifx\csname urlprefix\endcsname\relax\def\urlprefix{URL }\fi
\providecommand{\bibinfo}[2]{#2}
\providecommand{\eprint}[2][]{\url{#2}}

\bibitem[{\citenamefont{Pikovsky et~al.}(2001)\citenamefont{Pikovsky,
  Rosenblum, and Kurths}}]{Pikovsky:book}
\bibinfo{author}{\bibfnamefont{A.}~\bibnamefont{Pikovsky}},
  \bibinfo{author}{\bibfnamefont{M.}~\bibnamefont{Rosenblum}},
  \bibnamefont{and} \bibinfo{author}{\bibfnamefont{J.}~\bibnamefont{Kurths}},
  \emph{\bibinfo{title}{Synchronization, a universal concept in nonlinear
  sciences}} (\bibinfo{publisher}{Cambridge University Press},
  \bibinfo{address}{Cambridge}, \bibinfo{year}{2001}).

\bibitem[{\citenamefont{Schuster and Just}(2005)}]{Schuster:2005}
\bibinfo{author}{\bibfnamefont{H.~G.} \bibnamefont{Schuster}} \bibnamefont{and}
  \bibinfo{author}{\bibfnamefont{W.}~\bibnamefont{Just}},
  \emph{\bibinfo{title}{Deterministic Chaos}} (\bibinfo{publisher}{Wiley VCH},
  \bibinfo{address}{Weinheim}, \bibinfo{year}{2005}).

\bibitem[{\citenamefont{Pecora and Carroll}(1990)}]{Pecora:1990}
\bibinfo{author}{\bibfnamefont{L.~M.} \bibnamefont{Pecora}} \bibnamefont{and}
  \bibinfo{author}{\bibfnamefont{T.~L.} \bibnamefont{Carroll}},
  \bibinfo{journal}{Phys. Rev. Lett.} \textbf{\bibinfo{volume}{64}},
  \bibinfo{pages}{821} (\bibinfo{year}{1990}).

\bibitem[{\citenamefont{Cuomo and Oppenheim}(1993)}]{Cuomo:1993}
\bibinfo{author}{\bibfnamefont{K.~M.} \bibnamefont{Cuomo}} \bibnamefont{and}
  \bibinfo{author}{\bibfnamefont{A.~V.} \bibnamefont{Oppenheim}},
  \bibinfo{journal}{Phys. Rev. Lett.} \textbf{\bibinfo{volume}{71}},
  \bibinfo{pages}{65} (\bibinfo{year}{1993}).

\bibitem[{\citenamefont{Argyris et~al.}(2005)\citenamefont{Argyris, Syvridis,
  Larger, Annovazzi-Lodi, Colet, Fischer, Garc{\'i}a-Ojalvo, Mirasso, Pesquera,
  and Shore}}]{Argyris:2005}
\bibinfo{author}{\bibfnamefont{A.}~\bibnamefont{Argyris}},
  \bibinfo{author}{\bibfnamefont{D.}~\bibnamefont{Syvridis}},
  \bibinfo{author}{\bibfnamefont{L.}~\bibnamefont{Larger}},
  \bibinfo{author}{\bibfnamefont{V.}~\bibnamefont{Annovazzi-Lodi}},
  \bibinfo{author}{\bibfnamefont{P.}~\bibnamefont{Colet}},
  \bibinfo{author}{\bibfnamefont{I.}~\bibnamefont{Fischer}},
  \bibinfo{author}{\bibfnamefont{J.}~\bibnamefont{Garc{\'i}a-Ojalvo}},
  \bibinfo{author}{\bibfnamefont{C.~R.} \bibnamefont{Mirasso}},
  \bibinfo{author}{\bibfnamefont{L.}~\bibnamefont{Pesquera}}, \bibnamefont{and}
  \bibinfo{author}{\bibfnamefont{K.~A.} \bibnamefont{Shore}},
  \bibinfo{journal}{Nature} \textbf{\bibinfo{volume}{438}},
  \bibinfo{pages}{343} (\bibinfo{year}{2005}).

\bibitem[{\citenamefont{Klein et~al.}(2006{\natexlab{a}})\citenamefont{Klein,
  Gross, Kopelowitz, Rosenbluh, Khaykovich, Kinzel, and Kanter}}]{Klein:2006}
\bibinfo{author}{\bibfnamefont{E.}~\bibnamefont{Klein}},
  \bibinfo{author}{\bibfnamefont{N.}~\bibnamefont{Gross}},
  \bibinfo{author}{\bibfnamefont{E.}~\bibnamefont{Kopelowitz}},
  \bibinfo{author}{\bibfnamefont{M.}~\bibnamefont{Rosenbluh}},
  \bibinfo{author}{\bibfnamefont{L.}~\bibnamefont{Khaykovich}},
  \bibinfo{author}{\bibfnamefont{W.}~\bibnamefont{Kinzel}}, \bibnamefont{and}
  \bibinfo{author}{\bibfnamefont{I.}~\bibnamefont{Kanter}},
  \bibinfo{journal}{Phys. Rev. E} \textbf{\bibinfo{volume}{74}},
  \bibinfo{pages}{046201} (\bibinfo{year}{2006}{\natexlab{a}}).

\bibitem[{\citenamefont{Klein et~al.}(2005)\citenamefont{Klein, Mislovaty,
  Kanter, and Kinzel}}]{Klein:2005}
\bibinfo{author}{\bibfnamefont{E.}~\bibnamefont{Klein}},
  \bibinfo{author}{\bibfnamefont{R.}~\bibnamefont{Mislovaty}},
  \bibinfo{author}{\bibfnamefont{I.}~\bibnamefont{Kanter}}, \bibnamefont{and}
  \bibinfo{author}{\bibfnamefont{W.}~\bibnamefont{Kinzel}},
  \bibinfo{journal}{Phys. Rev. E} \textbf{\bibinfo{volume}{72}},
  \bibinfo{pages}{016214} (\bibinfo{year}{2005}).

\bibitem[{\citenamefont{Kanter et~al.}(2007)\citenamefont{Kanter, Gross, Klein,
  Kopelowitz, Yoskovits, Khaykovich, Kinzel, and
  Rosenbluh}}]{Kanter:Shutter:2007:PRL}
\bibinfo{author}{\bibfnamefont{I.}~\bibnamefont{Kanter}},
  \bibinfo{author}{\bibfnamefont{N.}~\bibnamefont{Gross}},
  \bibinfo{author}{\bibfnamefont{E.}~\bibnamefont{Klein}},
  \bibinfo{author}{\bibfnamefont{E.}~\bibnamefont{Kopelowitz}},
  \bibinfo{author}{\bibfnamefont{P.}~\bibnamefont{Yoskovits}},
  \bibinfo{author}{\bibfnamefont{L.}~\bibnamefont{Khaykovich}},
  \bibinfo{author}{\bibfnamefont{W.}~\bibnamefont{Kinzel}}, \bibnamefont{and}
  \bibinfo{author}{\bibfnamefont{M.}~\bibnamefont{Rosenbluh}},
  \bibinfo{journal}{Phys.\ Rev.\ Lett.} \textbf{\bibinfo{volume}{98}},
  \bibinfo{eid}{154101} (\bibinfo{year}{2007}).

\bibitem[{\citenamefont{Klein et~al.}(2006{\natexlab{b}})\citenamefont{Klein,
  Gross, Rosenbluh, Kinzel, Khaykovich, and Kanter}}]{Klein:06:PRE73}
\bibinfo{author}{\bibfnamefont{E.}~\bibnamefont{Klein}},
  \bibinfo{author}{\bibfnamefont{N.}~\bibnamefont{Gross}},
  \bibinfo{author}{\bibfnamefont{M.}~\bibnamefont{Rosenbluh}},
  \bibinfo{author}{\bibfnamefont{W.}~\bibnamefont{Kinzel}},
  \bibinfo{author}{\bibfnamefont{L.}~\bibnamefont{Khaykovich}},
  \bibnamefont{and} \bibinfo{author}{\bibfnamefont{I.}~\bibnamefont{Kanter}},
  \bibinfo{journal}{Phys. Rev. E} \textbf{\bibinfo{volume}{73}},
  \bibinfo{pages}{066214} (\bibinfo{year}{2006}{\natexlab{b}}).

\bibitem[{\citenamefont{Fischer et~al.}(2006)\citenamefont{Fischer, Vicente,
  Buldu, Peil, Mirasso, Torrent, and Garc{\'i}a-Ojalvo}}]{Fischer:2006:PRL}
\bibinfo{author}{\bibfnamefont{I.}~\bibnamefont{Fischer}},
  \bibinfo{author}{\bibfnamefont{R.}~\bibnamefont{Vicente}},
  \bibinfo{author}{\bibfnamefont{J.~M.} \bibnamefont{Buldu}},
  \bibinfo{author}{\bibfnamefont{M.}~\bibnamefont{Peil}},
  \bibinfo{author}{\bibfnamefont{C.~R.} \bibnamefont{Mirasso}},
  \bibinfo{author}{\bibfnamefont{M.~C.} \bibnamefont{Torrent}},
  \bibnamefont{and}
  \bibinfo{author}{\bibfnamefont{J.}~\bibnamefont{Garc{\'i}a-Ojalvo}},
  \bibinfo{journal}{Phys.\ Rev.\ Lett.} \textbf{\bibinfo{volume}{97}},
  \bibinfo{eid}{123902} (\bibinfo{year}{2006}).

\bibitem[{\citenamefont{Sivaprakasam et~al.}(2003)\citenamefont{Sivaprakasam,
  Paul, Spencer, Rees, and Shore}}]{Sivaprakasam:2003}
\bibinfo{author}{\bibfnamefont{S.}~\bibnamefont{Sivaprakasam}},
  \bibinfo{author}{\bibfnamefont{J.}~\bibnamefont{Paul}},
  \bibinfo{author}{\bibfnamefont{P.~S.} \bibnamefont{Spencer}},
  \bibinfo{author}{\bibfnamefont{P.}~\bibnamefont{Rees}}, \bibnamefont{and}
  \bibinfo{author}{\bibfnamefont{K.~A.} \bibnamefont{Shore}},
  \bibinfo{journal}{Opt. Lett.} \textbf{\bibinfo{volume}{28}},
  \bibinfo{pages}{1397} (\bibinfo{year}{2003}).

\bibitem[{\citenamefont{Lee et~al.}(2006)\citenamefont{Lee, Paul, Masoller, and
  Shore}}]{Lee:2006:JOSAB}
\bibinfo{author}{\bibfnamefont{M.~W.} \bibnamefont{Lee}},
  \bibinfo{author}{\bibfnamefont{J.}~\bibnamefont{Paul}},
  \bibinfo{author}{\bibfnamefont{C.}~\bibnamefont{Masoller}}, \bibnamefont{and}
  \bibinfo{author}{\bibfnamefont{K.~A.} \bibnamefont{Shore}},
  \bibinfo{journal}{J. Opt. Soc. Am. B} \textbf{\bibinfo{volume}{23}},
  \bibinfo{pages}{846} (\bibinfo{year}{2006}).

\bibitem[{\citenamefont{Cho}(2006)}]{Adrian:2006:SCI}
\bibinfo{author}{\bibfnamefont{A.}~\bibnamefont{Cho}},
  \bibinfo{journal}{Science} \textbf{\bibinfo{volume}{314}},
  \bibinfo{pages}{37} (\bibinfo{year}{2006}).

\bibitem[{\citenamefont{Engel et~al.}(1991)\citenamefont{Engel, K\"onig,
  Kreiter, and Singer}}]{Engel:1991:SCI}
\bibinfo{author}{\bibfnamefont{A.~K.} \bibnamefont{Engel}},
  \bibinfo{author}{\bibfnamefont{P.}~\bibnamefont{K\"onig}},
  \bibinfo{author}{\bibfnamefont{A.~K.} \bibnamefont{Kreiter}},
  \bibnamefont{and} \bibinfo{author}{\bibfnamefont{W.}~\bibnamefont{Singer}},
  \bibinfo{journal}{Science} \textbf{\bibinfo{volume}{252}},
  \bibinfo{pages}{1177} (\bibinfo{year}{1991}).

\bibitem[{\citenamefont{Campbell and Wang}(1998)}]{Campbell:1998}
\bibinfo{author}{\bibfnamefont{S.~R.} \bibnamefont{Campbell}} \bibnamefont{and}
  \bibinfo{author}{\bibfnamefont{D.}~\bibnamefont{Wang}},
  \bibinfo{journal}{Physica D} \textbf{\bibinfo{volume}{111}},
  \bibinfo{pages}{151} (\bibinfo{year}{1998}).

\bibitem[{\citenamefont{Rosenbluh et~al.}(2007)\citenamefont{Rosenbluh, Aviad,
  Cohen, Khaykovich, Kinzel, Kopelowitz, Yoskovits, and
  Kanter}}]{Rosenbluh:SpikingOpticalPatterns:2007:PRE}
\bibinfo{author}{\bibfnamefont{M.}~\bibnamefont{Rosenbluh}},
  \bibinfo{author}{\bibfnamefont{Y.}~\bibnamefont{Aviad}},
  \bibinfo{author}{\bibfnamefont{E.}~\bibnamefont{Cohen}},
  \bibinfo{author}{\bibfnamefont{L.}~\bibnamefont{Khaykovich}},
  \bibinfo{author}{\bibfnamefont{W.}~\bibnamefont{Kinzel}},
  \bibinfo{author}{\bibfnamefont{E.}~\bibnamefont{Kopelowitz}},
  \bibinfo{author}{\bibfnamefont{P.}~\bibnamefont{Yoskovits}},
  \bibnamefont{and} \bibinfo{author}{\bibfnamefont{I.}~\bibnamefont{Kanter}},
  \bibinfo{journal}{Phys.\ Rev.\ E} \textbf{\bibinfo{volume}{76}},
  \bibinfo{eid}{046207} (\bibinfo{year}{2007}).

\bibitem[{\citenamefont{Atay et~al.}(2004)\citenamefont{Atay, Jost, and
  Wende}}]{Atay:2004:PRL}
\bibinfo{author}{\bibfnamefont{F.~M.} \bibnamefont{Atay}},
  \bibinfo{author}{\bibfnamefont{J.}~\bibnamefont{Jost}}, \bibnamefont{and}
  \bibinfo{author}{\bibfnamefont{A.}~\bibnamefont{Wende}},
  \bibinfo{journal}{Phys. Rev. Lett.} \textbf{\bibinfo{volume}{92}},
  \bibinfo{eid}{144101} (\bibinfo{year}{2004}).

\bibitem[{\citenamefont{Matskiv et~al.}(2004)\citenamefont{Matskiv, Maistrenko,
  and Mosekilde}}]{Matskiv:2004}
\bibinfo{author}{\bibfnamefont{I.}~\bibnamefont{Matskiv}},
  \bibinfo{author}{\bibfnamefont{Y.}~\bibnamefont{Maistrenko}},
  \bibnamefont{and}
  \bibinfo{author}{\bibfnamefont{E.}~\bibnamefont{Mosekilde}},
  \bibinfo{journal}{Physica D} \textbf{\bibinfo{volume}{199}},
  \bibinfo{pages}{45} (\bibinfo{year}{2004}).

\bibitem[{\citenamefont{Masoller and Mart\'{\i}}(2005)}]{Masoller:2005}
\bibinfo{author}{\bibfnamefont{C.}~\bibnamefont{Masoller}} \bibnamefont{and}
  \bibinfo{author}{\bibfnamefont{A.~C.} \bibnamefont{Mart\'{\i}}},
  \bibinfo{journal}{Phys. Rev. Lett.} \textbf{\bibinfo{volume}{94}},
  \bibinfo{eid}{134102} (pages~\bibinfo{numpages}{4}) (\bibinfo{year}{2005}).

\bibitem[{\citenamefont{Topaj et~al.}(2001)\citenamefont{Topaj, Kye, and
  Pikovsky}}]{Topaj:2001}
\bibinfo{author}{\bibfnamefont{D.}~\bibnamefont{Topaj}},
  \bibinfo{author}{\bibfnamefont{W.-H.} \bibnamefont{Kye}}, \bibnamefont{and}
  \bibinfo{author}{\bibfnamefont{A.}~\bibnamefont{Pikovsky}},
  \bibinfo{journal}{Phys. Rev. Lett.} \textbf{\bibinfo{volume}{87}},
  \bibinfo{pages}{074101} (\bibinfo{year}{2001}).

\bibitem[{\citenamefont{Kestler et~al.}(2007)\citenamefont{Kestler, Kinzel, and
  Kanter}}]{Sublattice:2007:PRE}
\bibinfo{author}{\bibfnamefont{J.}~\bibnamefont{Kestler}},
  \bibinfo{author}{\bibfnamefont{W.}~\bibnamefont{Kinzel}}, \bibnamefont{and}
  \bibinfo{author}{\bibfnamefont{I.}~\bibnamefont{Kanter}},
  \bibinfo{journal}{Phys. Rev. E} \textbf{\bibinfo{volume}{76}},
  \bibinfo{eid}{035202} (pages~\bibinfo{numpages}{4}) (\bibinfo{year}{2007}).

\bibitem[{\citenamefont{Golubitsky and Stewart}(2006)}]{Golubitsky:2006}
\bibinfo{author}{\bibfnamefont{M.}~\bibnamefont{Golubitsky}} \bibnamefont{and}
  \bibinfo{author}{\bibfnamefont{I.}~\bibnamefont{Stewart}},
  \bibinfo{journal}{Bull. Amer. Math. Soc.} \textbf{\bibinfo{volume}{43}},
  \bibinfo{pages}{305} (\bibinfo{year}{2006}).

\bibitem[{\citenamefont{Lepri et~al.}(1993)\citenamefont{Lepri, Giacomelli,
  Politi, and Arecchi}}]{Lepri:1993:PHD}
\bibinfo{author}{\bibfnamefont{S.}~\bibnamefont{Lepri}},
  \bibinfo{author}{\bibfnamefont{G.}~\bibnamefont{Giacomelli}},
  \bibinfo{author}{\bibfnamefont{A.}~\bibnamefont{Politi}}, \bibnamefont{and}
  \bibinfo{author}{\bibfnamefont{F.~T.} \bibnamefont{Arecchi}},
  \bibinfo{journal}{Physica D} \textbf{\bibinfo{volume}{70}},
  \bibinfo{pages}{235} (\bibinfo{year}{1993}).

\bibitem[{\citenamefont{Lang and Kobayashi}(1980)}]{Lang:1980}
\bibinfo{author}{\bibfnamefont{R.}~\bibnamefont{Lang}} \bibnamefont{and}
  \bibinfo{author}{\bibfnamefont{K.}~\bibnamefont{Kobayashi}},
  \bibinfo{journal}{IEEE J.\ Quantum Electron.} \textbf{\bibinfo{volume}{16}},
  \bibinfo{pages}{347} (\bibinfo{year}{1980}).

\bibitem[{\citenamefont{Ahlers et~al.}(1998)\citenamefont{Ahlers, Parlitz, and
  Lauterborn}}]{Ahlers:1998}
\bibinfo{author}{\bibfnamefont{V.}~\bibnamefont{Ahlers}},
  \bibinfo{author}{\bibfnamefont{U.}~\bibnamefont{Parlitz}}, \bibnamefont{and}
  \bibinfo{author}{\bibfnamefont{W.}~\bibnamefont{Lauterborn}},
  \bibinfo{journal}{Phys.\ Rev.\ E} \textbf{\bibinfo{volume}{58}},
  \bibinfo{pages}{7208} (\bibinfo{year}{1998}).

\bibitem[{\citenamefont{Abarbanel et~al.}(1996)\citenamefont{Abarbanel, Rulkov,
  and Sushchik}}]{Abarbanel:1996:PRE}
\bibinfo{author}{\bibfnamefont{H.~D.~I.} \bibnamefont{Abarbanel}},
  \bibinfo{author}{\bibfnamefont{N.~F.} \bibnamefont{Rulkov}},
  \bibnamefont{and} \bibinfo{author}{\bibfnamefont{M.~M.}
  \bibnamefont{Sushchik}}, \bibinfo{journal}{Phys. Rev. E}
  \textbf{\bibinfo{volume}{53}}, \bibinfo{pages}{4528} (\bibinfo{year}{1996}).

\bibitem[{\citenamefont{Boccaletti et~al.}(2002)\citenamefont{Boccaletti,
  Kurths, Osipov, Valladares, and Zhou}}]{Boccaletti:2002:PRep}
\bibinfo{author}{\bibfnamefont{S.}~\bibnamefont{Boccaletti}},
  \bibinfo{author}{\bibfnamefont{J.}~\bibnamefont{Kurths}},
  \bibinfo{author}{\bibfnamefont{G.}~\bibnamefont{Osipov}},
  \bibinfo{author}{\bibfnamefont{D.~L.} \bibnamefont{Valladares}},
  \bibnamefont{and} \bibinfo{author}{\bibfnamefont{C.~S.} \bibnamefont{Zhou}},
  \bibinfo{journal}{Phys. Rep.} \textbf{\bibinfo{volume}{366}},
  \bibinfo{pages}{1} (\bibinfo{year}{2002}).

\end{thebibliography}

\end{document}